# Pressure-Induced Decomposition of β-SnWO$_4$

Sergio Ferrari[1,2], Daniel Diaz-Anichtchenko[3], Pablo Botella[4], Jordi Ibáñez[5], Robert Oliva[5], Alexei Kuzmin[6], Alfonso Muñoz[7], Frederico Alabarse[8], Daniel Errandonea[4,*]

[1]Departamento Física Experimental, CNEA, Centro Atómico Constituyentes, Av. Gral. Paz 1499, San Martín 1650, Argentina

[2]Consejo Nacional de Ciencia y Tecnología, Ciudad Autónoma de Buenos Aires 1650, Argentina

[3]Universidad Europea de Valencia, Valencia, Comunidad Valenciana, Spain

[4]Departamento de Física Aplicada-ICMUV, MALTA Consolider Team, Universidad de Valencia, Dr. Moliner 50, Burjassot, 46100, Valencia, Spain

[5]Geosciences Barcelona (GEO3BCN-CSIC), MALTA Consolider Team, Barcelona, 08028, Spain

[6]Institute of Solid State Physics, University of Latvia, Kengaraga street 8, Riga LV-1063, Latvia

[7]Departamento de Física, MALTA-Consolider Team, Universidad de La Laguna, San Cristobal de La Laguna, E-38200, Tenerife, Spain

[8]Elettra Sincrotrone Trieste, Trieste, 34149, Italy

*Corresponding author; email: daniel.errandonea@uv.es; ORCID ID: 0000-0003-0189-4221

Keywords: SnWO$_4$, High pressure, XRD, Decomposition, DFT, Equation of state

**Abstract**

This study reports the decomposition of β-SnWO$_4$ into Sn, SnO$_2$, and WO$_3$ induced by static compression. We performed high-pressure synchrotron powder angle-dispersive X-ray diffraction measurements and found that decomposition occurs at a pressure of 13.97(5) GPa and is irreversible. This result contradicts a previous study that, based on density-functional theory calculations and crystal-chemistry arguments, predicted a pressure-driven transition from β-SnWO$_4$ to α-SnWO$_4$. Our analysis indicates that the observed decomposition is unrelated to mechanical or dynamic instabilities. Instead, it likely stems from frustration of the β→α transition, as this transformation requires a change in Sn coordination from octahedral to tetrahedral. The assessment of how pressure influences the volume of the unit cell provided an accurate determination of the room-temperature pressure–volume equation of state for β-SnWO$_4$. Furthermore, the elastic constants and moduli, as well as the pressure dependence of Raman and infrared modes of β-SnWO$_4$, were derived from density-functional theory calculations. Several phonon modes exhibited softening, and three cases of phonon anti-crossing were observed.

## 1. Introduction

SnWO$_4$ is a promising catalyst for photoelectrochemical water splitting due to its narrow band gap and appropriate band alignment [1]. Furthermore, SnWO$_4$ has great potential as a viable anode material for lithium-ion batteries [2]. Several studies highlight additional applications of tin tungstate, such as its use in transparent conducting oxides [3] and gas sensors [4]. Additionally, the polyhedral distortions present in the Sn$^{2+}$ environment result in a metallic off-center position, which has been extensively researched for its potential application in multiferroic materials [5]. SnWO$_4$ has been widely studied not only for its technological applications but also for its interest in fundamental research. This compound is unique among other orthotungstates due to the influence of the 5s$^2$ lone electron pair (LEP) associated with divalent tin on its crystal structure. SnWO$_4$ displays two distinctly different structural configurations which were solved by Jeitschko and Sleight [6,7]. They first determined the crystal structure of β-SnWO$_4$ [6] and subsequently the structure of α-SnWO$_4$ [7], which are represented in Figures 1(a) and 1(b), respectively.

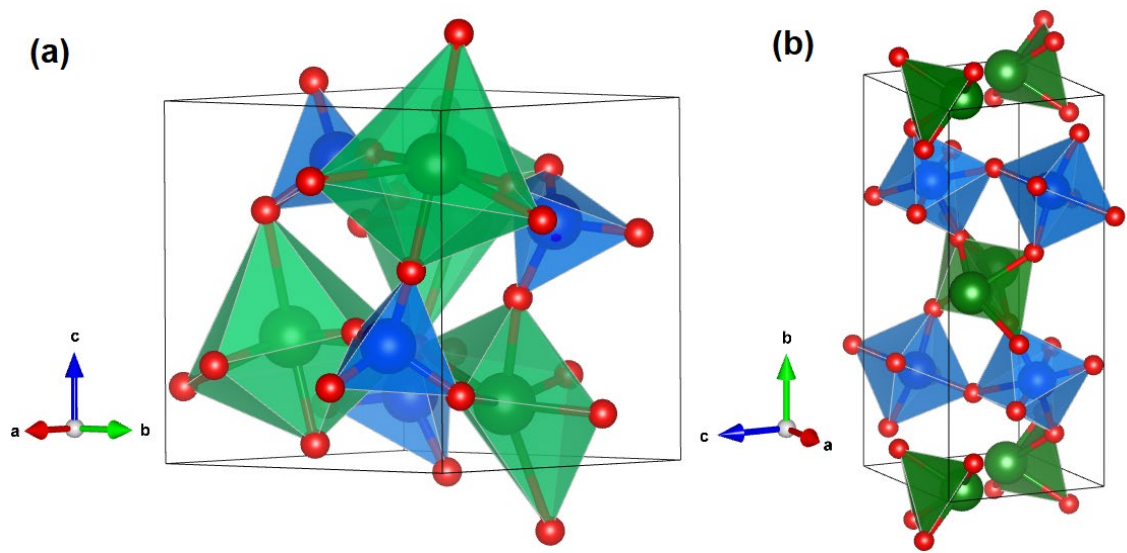

**Figure 1:** Crystal structure of (a) β-SnWO$_4$ and (b) α-SnWO$_4$. The coordination polyhedra of Sn (W) are shown in green (blue). Oxygen atoms are represented in red.

β-SnWO$_4$ is cubic and described by space group $P2_13$. The W atom is situated within a relatively regular tetrahedral arrangement of oxygen atoms. Sn atoms are coordinated to six oxygen atoms, resulting in the formation of three short bonds and three long bonds (see Figure 1(a)), which is indicative of a cation with a LEP configuration. α-SnWO$_4$ is orthorhombic and described by space group *Pnna*. The W atom is coordinated in an octahedral arrangement by oxygen atoms, with the WO$_6$ octahedra interconnected at four corners. This configuration results in the formation of sheets of [WO$_4$]$^{2-}$ polyanions, which are stabilized by Sn$^{2+}$. Tin atoms are coordinated by four oxygen atoms, forming a trigonal bipyramidal arrangement, with the Sn atom positioned at the apex of the pyramid (see Figure 1(b)). α-SnWO$_4$ is the stable phase at ambient conditions. It undergoes a reversible phase transition to β-SnWO$_4$ at 670 °C [6]. This phase can be quenched to ambient conditions upon rapid cooling.

It has been proposed, based on density-functional theory (DFT) calculations, that β-SnWO$_4$ should transform into α-SnWO$_4$ under the application of hydrostatic pressure [8,9]. High pressure (HP) serves as a highly effective method for examining materials. The application of pressure facilitates substantial changes in interatomic distances, which, in turn, permits the modification of essential physical characteristics [10]. This technique is particularly advantageous for the analysis of compounds containing atoms with a lone electron pair, like SnWO$_4$, leading to structures characterized by extensive voids and the possibility of polyhedral tilting [11]. α-SnWO$_4$ has been experimentally studied under compression [12]. Two phase transitions were found, one at 12.9 GPa and the other at 17.5 GPa [12]. The phase transition results in an increase in the coordination number of Sn atoms. The alteration in the coordination of Sn atoms arises from the suppression of the LEP and the establishment of additional bonds with second neighboring oxygen atoms. In contrast to α-SnWO$_4$, β-SnWO$_4$ has not been studied experimentally under HP conditions. Here, we report a powder X-ray diffraction (XRD) study of β-SnWO$_4$ under HP, which is supported by DFT calculations. Our study provides undoubtful evidence that β-SnWO$_4$ undergoes chemical decomposition instead of a structural phase transition. The pressure-volume (P-V) equation of state (EOS) and the elastic constants of β-SnWO$_4$ will also be presented, as well as the calculated pressure dependence of Raman and infrared (IR) modes. We find that several phonons undergo softening, while others show anti-crossing behavior.

## 2. Methods

*2.1 Experiments*

Polycrystalline β-SnWO$_4$ was prepared following the method described previously by Jeitschko and Sleight [6]. Equimolar amounts of SnO (99.99%) and WO$_3$ (99.9%) powders were mechanically mixed and sealed in a silica ampoule under a vacuum. The ampoule was heated at 800 °C for 8 h and then rapidly quenched to room temperature.

Experiments were performed using a membrane-driven diamond-anvil cell (DAC) with 500 μm culet diamonds. A stainless-steel gasket, pre-indented to a thickness of 50 μm and containing a centered hole with a diameter of 150 μm, was employed to contain a fine powder obtained from the synthesized material. The pressure-transmitting medium used was a mixture of ethanol, methanol, and water in a ratio of 16:3:1, which enables quasi-hydrostatic conditions at pressures up to 10 GPa [13]. In oxides, this method has been effectively used for precise studies at pressures like those covered by this study [14]. Copper powder was loaded together with the sample as an internal standard for pressure determination. We used for calibration the EOS of Cu reported by Dewaele *et al*. [15]. The pressure measurements exhibited an error margin of less than 0.05 GPa. Angle-resolved HP-XRD measurements were performed at the Xpress beamline of the Elettra Synchrotron Radiation Facility, employing a monochromatic wavelength of 0.4956 Å and a PILATUS 3S 6M detector. Calibration of the instrument was achieved using LaB$_6$. The XRD patterns were generated by integrating the two-dimensional diffraction rings captured by the detector with the aid of Dioptas [16]. The X-ray beam was focused to a spot size of 50 μm × 50 μm. The analysis of XRD patterns was carried out using MAUD [17].

*2.2 Density functional theory*

*Ab initio* calculations were performed utilizing DFT [18] and the projector-augmented wave method [19] within the Vienna *ab initio* simulation package (VASP) [20]. An energy cutoff of 560 eV was set to ensure accurate results. The exchange-correlation energy was characterized by using the generalized-gradient approximation with the Perdew-Burke-Ernzerhof for solids functional [21]. The Monkhorst-Pack method [22] was applied to sample the Brillouin zone (BZ) integrations, employing a 4 x 4 x 4 mesh in the irreducible BZ. In the optimized equilibrium state, the forces acting on the atoms were maintained below 1 meV/Å along the Cartesian axes. For the computation of the vibrational properties, we used the Phonopy package [23]. A 2 x 2 x 2 supercell was employed to derive the phonon dispersion and to assess dynamical stability. The mechanical properties were examined by calculating the elastic constants through stress-strain analyses conducted in VASP using the Le Page method [24]. The different elastic moduli were subsequently derived from these elastic constants.

## 3. Results and discussion

Figure 2(a) shows the XRD pattern we recorded at the lowest pressure measured, 0.43 GPa. The pattern can be explained by β-SnWO$_4$ as shown by the Rietveld refinement included in the figure. A 3% fraction of α-SnWO$_4$ was detected coexisting with β-SnWO$_4$ together with a small fraction of the precursors. However, weak peaks not belonging to β-SnWO$_4$ do not interfere with the peaks of the sample and therefore do not affect the study of its crystal structure. The unit-cell parameter of β-SnWO$_4$ at 0.43 GPa is 7.2361(5) Å. As pressure increased from 0.43 GPa to 13.03 GPa we did not find any evidence of a phase transition. As shown in Figure 3, we only observed a gradual shift of XRD peaks towards high angles due to the contraction of the unit-cell parameter. In Figure 2(b) we show the XRD pattern measured at 13.03 GPa including the Rietveld refinement. The quality of the refinement at 13.03 GPa is as good as for the refinement at 0.43 GPa. The unit-cell parameter at 13.03 GPa is 6.7366(5) Å.

The pressure dependence of the lattice parameter of β-SnWO$_4$ was extracted from the XRD patterns measured up to 13.03 GPa. The results are summarized in Table 1. In Figure 4 we represent the pressure dependence of the unit-cell volume compared with the results of our DFT calculations and results from previous experiments performed on α-SnWO$_4$ [12]. All these data show that β-SnWO$_4$ is much more compressible than α-SnWO$_4$. Also it can be seen that experiments and DFT agree quite well up to 8 GPa, with the compressibility obtained from calculations being larger than the experimental one about this pressure. The unit-cell volume calculated at 0 GPa is 388.84 Å$^3$ which is in excellent agreement with the unit-cell volume measured from XRD at ambient pressure, 388.84(9) Å$^3$ [6].

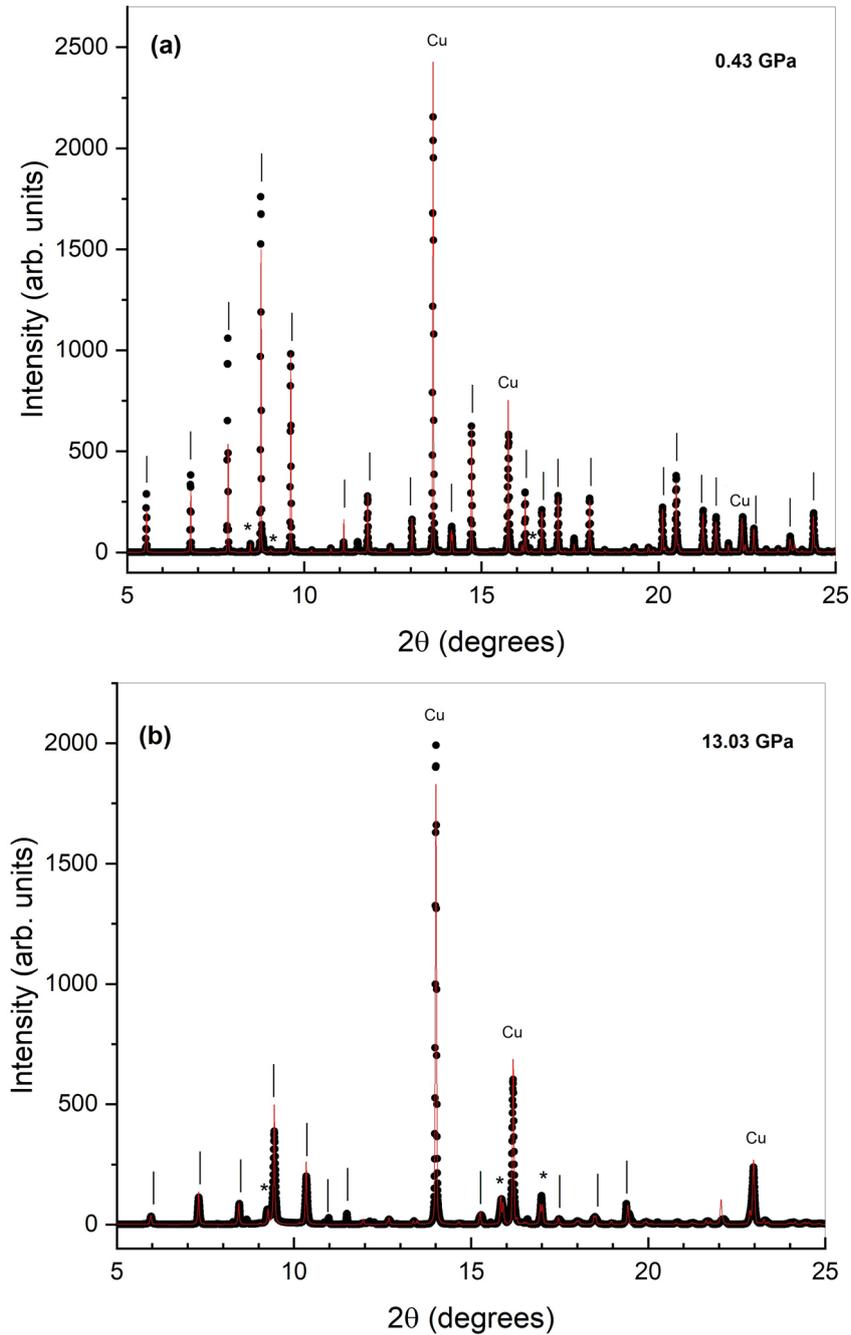

**Figure 2:** XRD pattern measured (λ=0.4956 Å) at 0.43 GPa (a) and 13.03 GPa (b). The dots are the experiments and the red lines are the refinements. The peaks from Cu, used to determine the pressure, are identified. Ticks indicate the position of the peaks from β-SnWO$_4$ and the most intense peaks from α-SnWO$_4$ are identified by asterisks.

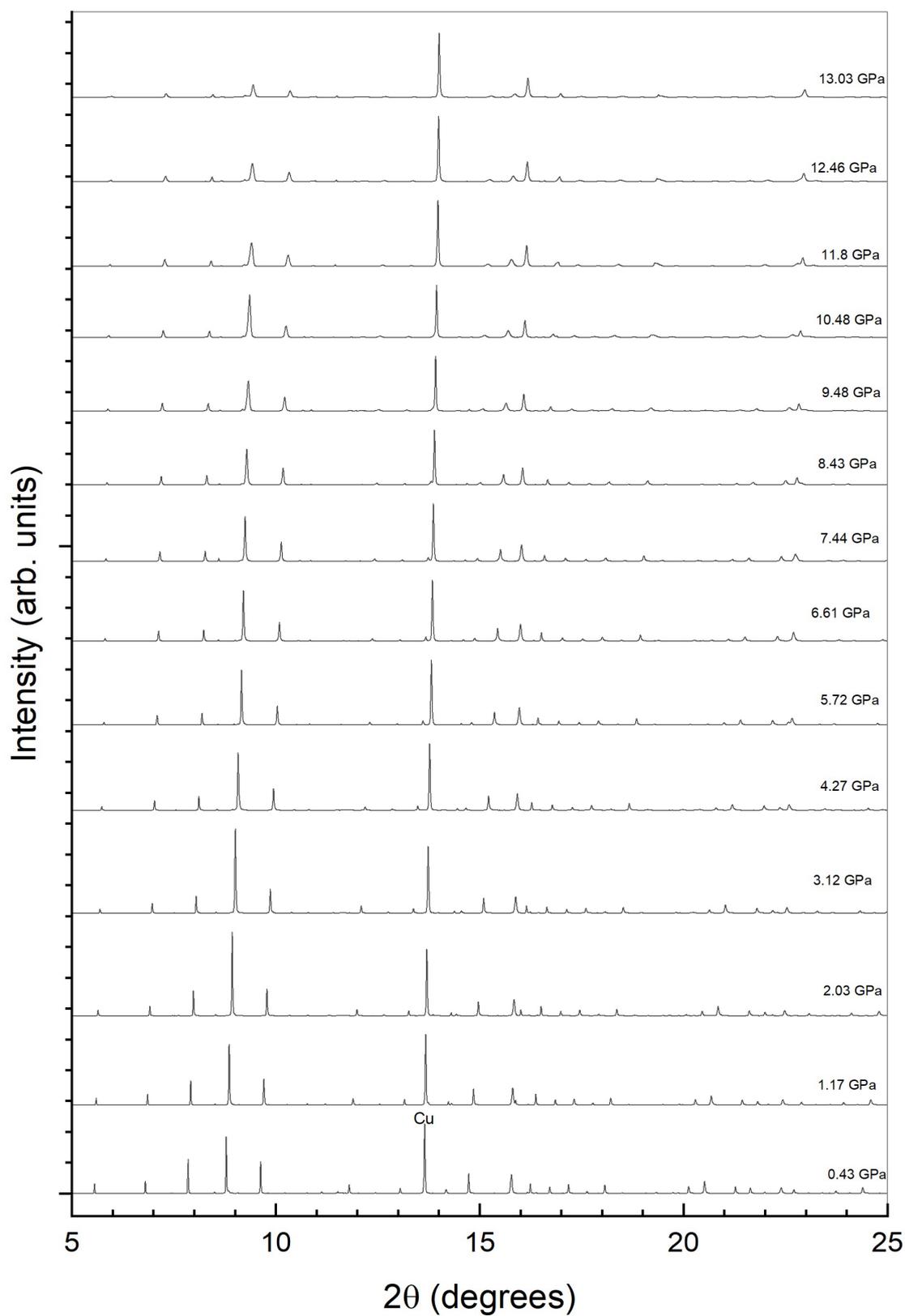

**Figure 3:** Selection of XRD patterns (λ=0.4956 Å) measured from 0.43 GPa to 13.03 GPa. The most intense peak of Cu is identified at the lowest pressure and can be easily followed at all pressures.

**Table 1:** Unit-cell parameter determined for β-SnWO$_4$ from XRD as a function of pressure.

| Pressure (GPa) | Unit-cell parameter (Å) | Pressure (GPa) | Unit-cell parameter (Å) |
|---|---|---|---|
| 0.43(5) | 7.2361(5) | 6.61(5) | 6.9082(5) |
| 0.79(5) | 7.2086(5) | 7.44(5) | 6.8804(5) |
| 1.17(5) | 7.1753(5) | 7.91(5) | 6.8601(5) |
| 1.62(5) | 7.1575(5) | 8.43(5) | 6.8545(5) |
| 2.03(5) | 7.1182(5) | 9.10(5) | 6.8365(5) |
| 2.48(5) | 7.1018(5) | 9.48(5) | 6.8288(5) |
| 3.12(5) | 7.0575(5) | 10.48(5) | 6.7988(5) |
| 3.82(5) | 7.0334(5) | 10.68(5) | 6.7936(5) |
| 4.27(5) | 7.0034(5) | 11.00(5) | 6.7901(5) |
| 4.74(5) | 6.9858(5) | 11.80(5) | 6.7685(5) |
| 5.72(5) | 6.9449(5) | 12.43(5) | 6.7416(5) |
| 5.90(5) | 6.9337(5) | 13.03(5) | 6.7366(5) |

The dependence of the volume on pressure has been examined utilizing a third-order Birch-Murnaghan EOS [25]. We have determined the volume at zero pressure, $V_0$ = 386.9(1.4) Å$^3$, the bulk modulus at zero pressure $B_0$ = 22.9(1.6) GPa, and its pressure derivative $B_0'$ = 7.7(3). The implied value of the second pressure derivative of the bulk modulus [26] is $B_0''$ = -0.7941 GPa$^{-1}$. From our DFT calculations we obtained $V_0$ = 387.6(4) Å$^3$, $B_0$ = 23.5(4) GPa, $B_0'$ = 5.8(1), and $B_0''$ = -0.3856 GPa$^{-1}$ (implied value). The values of $V_0$ and $B_0$ agree with experiments within one standard deviation. $B_0'$ is 18% smaller in calculations than in the experiment, which is a consequence of the smaller compressibility of DFT results above 8 GPa (see Figure 4). The EOS parameters of β-SnWO$_4$ are compared in Table 2 with those of α-SnWO$_4$. The bulk modulus of β-SnWO$_4$ is four times smaller than the bulk modulus of α-SnWO$_4$ (see Table 2). This conclusion is in line with results from calculations performed using the linear combination of atomic orbitals method and the code CRYSTAL09 [9]. According to these calculations, the bulk modulus of β-SnWO$_4$ is 13.75 GPa, which is smaller than the value obtained from our calculations and experiments (23.5 GPa).

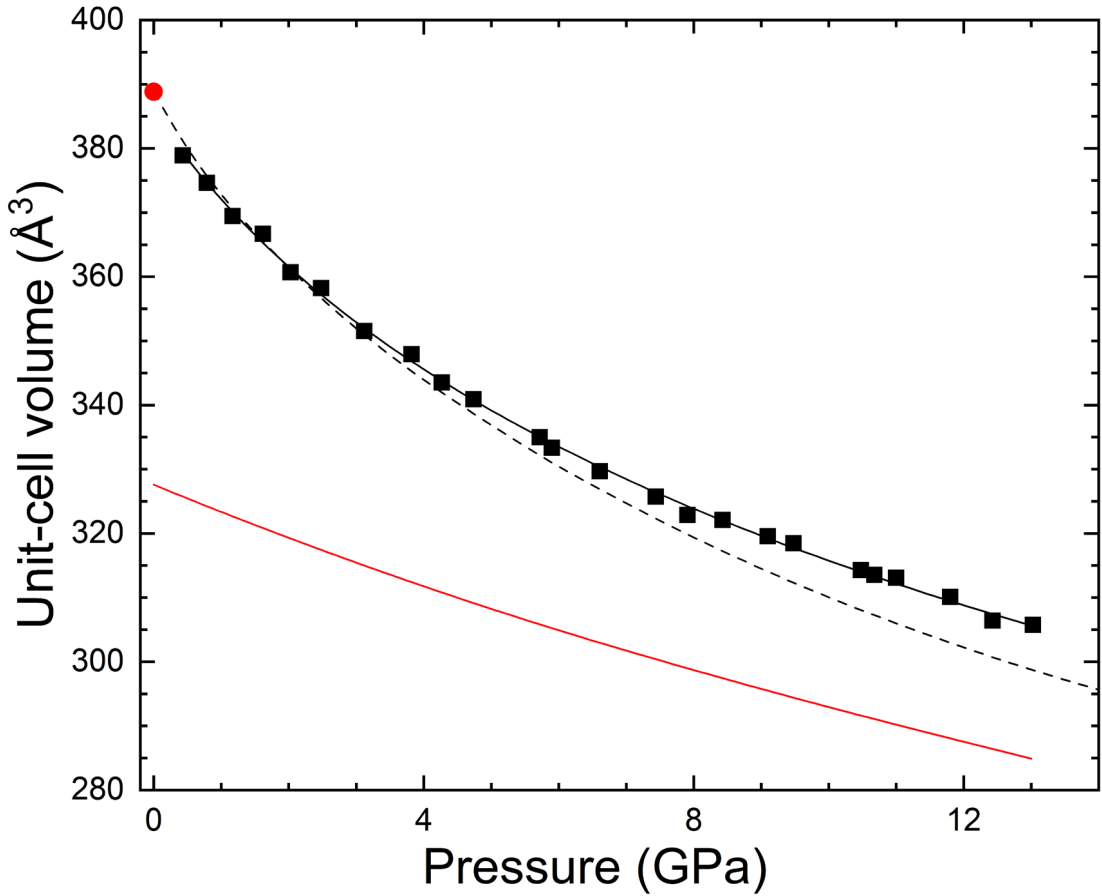

**Figure 4:** Pressure dependence of the unit-cell volume of β-SnWO$_4$. Squares are the results of present experiments. The circle is from ambient pressure studies [6]. The black solid line is the EOS determined from experiments. The black dashed line represents DFT results. The red solid line is the EOS determined previously for α-SnWO$_4$ from experiments [12].

**Table 2:** EOS parameters of β-SnWO$_4$ obtained from the present study. The results from previous experiments on α-SnWO$_4$ [12].

| Material | Method | $V_0$ (Å$^3$) | $B_0$ (GPa) | $B_0'$ |
|---|---|---|---|---|
| β-SnWO$_4$ | XRD This work | 386.9(1.4) | 22.9(1.6) | 7.7(3) |
| β-SnWO$_4$ | DFT This work | 387.6(4) | 23.5(4) | 5.8(1) |
| α-SnWO$_4$ | XRD [12] | 327.6 | 73.6(5.6) | 3.4(1.5) |
| α-SnWO$_4$ | DFT [12] | 327.6 | 87.9 | 4 |

The present results also indicate that the bulk modulus of β-SnWO$_4$ is much smaller than that of any AWO$_4$ orthotungstate studied until now. Scheelite-type compounds (BaWO$_4$, PbWO$_4$, EuWO$_4$, SrWO$_4$, and CaWO$_4$) have bulk moduli between 57 and 79 GPa [27]. Wolframite-type compounds (MgWO$_4$, MnWO$_4$, ZnWO$_4$, and CdWO$_4$) have bulk moduli between 123 and 160 GPa [28]. Thus, our results suggest that β-SnWO$_4$ is the most compressible orthotungstate. Since WO$_4$ and WO$_6$ polyhedra are incompressible, it is usually assumed that the bulk modulus of AWO$_4$ orthotungstates is determined by the compressibility of the coordination polyhedron of the divalent cation A [27]. In fact, it has been found that it is inversely proportional to the cube of the A–O bond distance [27]. Such a model gives a bulk modulus of 85 GPa for α-SnWO$_4$, which agrees with the known value (see Table 2), and of 78 GPa for β-SnWO$_4$, which is a largely overestimated value. This means that the contraction of the volume in β-SnWO$_4$ is not only due to the volume reduction of the SnO$_6$ octahedron. We think that the large compressibility (small bulk modulus) is related to the fact that β-SnWO$_4$ has a small density, with a larger "empty" space within the structure. Notice that the unit-cell of β-SnWO$_4$ is 18% larger than that in α-SnWO$_4$. The low density of β-SnWO$_4$ makes the electronic repulsion between cations much smaller in β-SnWO$_4$ than in any other AWO$_4$ orthotungstate. The volume occupied by SnO$_6$ and WO$_4$ units in β-SnWO$_4$ is 21% of the unit-cell volume, which is much smaller than in any other orthotungstate. This is the main reason behind the fact that β-SnWO$_4$ has a bulk modulus as small as that of metal iodates, the most compressible known oxides [11,29], and hybrid perovskites, compounds where compressibility is very high due to the organic molecules included in the crystal structure [30].

When increasing pressure from 13.03 GPa to 13.94 GPa we found considerable changes in the XRD pattern indicating a change in the structure. We illustrate it in Figure 5(a) with a pattern measured at 14.11 GPa. At this pressure the peaks of β-SnWO$_4$ broaden, and additional peaks appear that are sharper and more intense than those of β-SnWO$_4$. For instance, the extra peaks at 8.7° and 11.5° cannot be indexed by a single structure. Two of them correspond to the expected positions of the peaks of α-SnWO$_4$, but the remaining extra peaks cannot be assigned to it.

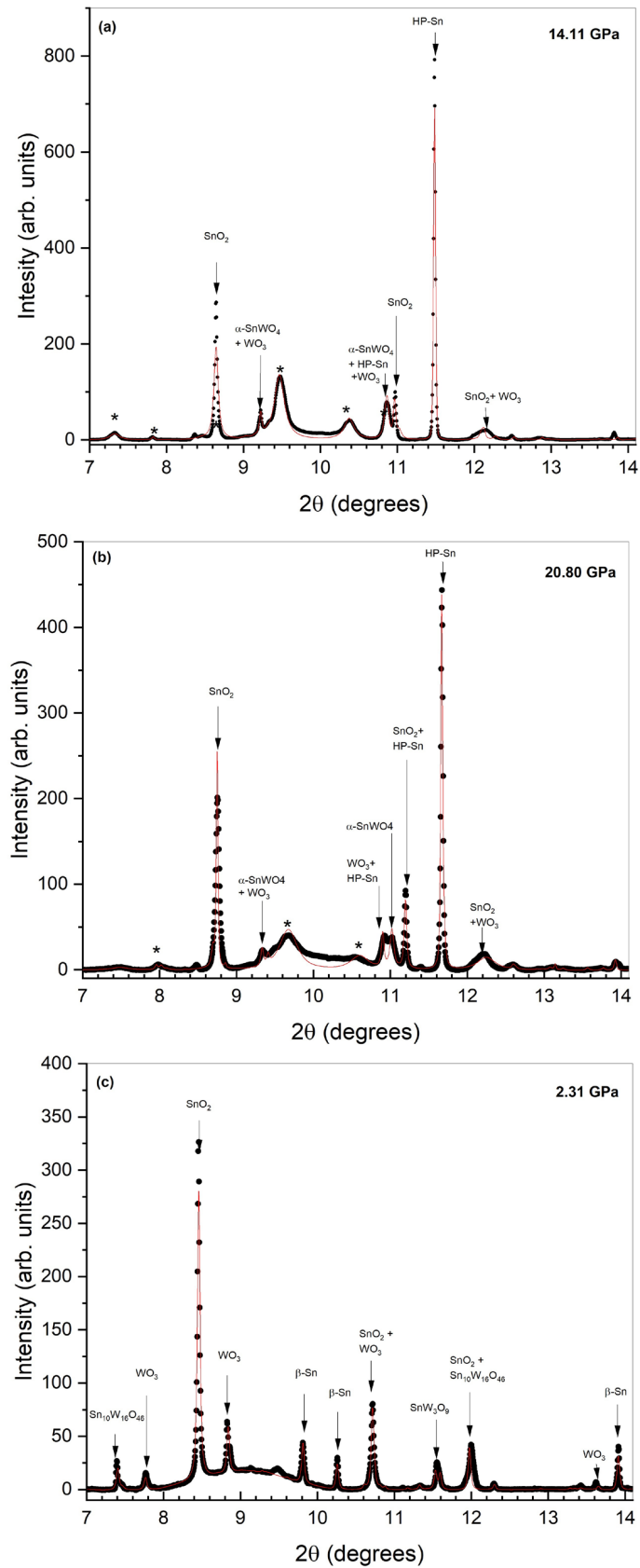

**Figure 5:** XRD pattern (λ=0.4956 Å) measured at 14.11 GPa (a) and 20.8 GPa (b) during compression, and 2.31 GPa after decompression (c). The dots are the experiments, and the red lines are the refinements. The peaks from different decomposition products are identified. In (a) and (b) the asterisks identify peaks from β-SnWO$_4$.

To explain the extra peaks in the XRD pattern we also considered the two high-pressure phases of α-SnWO$_4$ [12]. However, they are also unable to explain the diffraction patterns measured above 13 GPa. In contrast, we found they correspond to the peaks of SnO$_2$ [31] and WO$_3$ [32] at 14.11 GPa, and to peaks that could be assigned to HP phase of Sn (known as γ phase) [33]. Figure 5(a) shows the Rietvekd refinement assuming β-SnWO$_4$ and the decomposition products. All the most intese peaks can be explained assuming this hypothesis. This indicates that the transition to α-SnWO$_4$ was frustrated and instead β-SnWO$_4$ prefers to decompose via the reaction 2 SnWO$_4$ → Sn + SnO$_2$ + 2WO$_3$. A similar phenomenon was observed in thin films of SnWO$_4$ when the temperature of synthesis was 500–650 °C [34].

With increasing pressure, the β-SnWO$_4$ peaks broaden further and decrease in intensity, indicating progressive amorphization. This can be seen in Figure 5(b) which corresponds to a pressure of 20.80 GPa. Under decompression, the decomposition of β-SnWO$_4$ is irreversible, as shown in Figure 5(c). The XRD pattern measured at 2.31 GPa under decompression shows peaks that can be assigned to the low-pressure of Sn (β phase [33]), SnW$_3$O$_9$, Sn$_{10}$W$_{16}$O$_{46}$, SnO$_2$, and WO$_3$, along with a diffuse halo from 8° to 10°, which corresponds to an amorphous component that could be associated with the pressure medium [35]. Additionally, there are also a few extra peaks that could not be identified with any known Sn and W oxides or with any tin tungstate reported in the Inorganic Crystal Structure Database. This suggests that compounds with compositions different from those already known could form upon decompression. The irreversibility of this phenomenon supports our interpretation that pressure induces the chemical decomposition of β-SnWO$_4$ [36]. Our findings rule out previous predictions that a β→α phase transition should occur under compression in SnWO$_4$ [8].

We will now discuss the possible reasons for the reported decomposition of β-SnWO$_4$. First, we would like to remark that β-SnWO$_4$ is a metastable phase at room temperature, with α-SnWO$_4$ being the stable polymorph [12]. This agrees with the results of our DFT calculations. In Figure 6, we show the calculated enthalpies versus pressure for both phases and the decomposition here proposed (SnWO$_4$ → $\frac{1}{2}$Sn + $\frac{1}{2}$SnO$_2$ + WO$_3$). At ambient pressure, the enthalpy of β-SnWO$_4$ is 2 eV larger than that of α-SnWO$_4$ and 0.2 eV larger than the proposed decomposition. This implies that β-SnWO$_4$ is thermodynamically unstable, while α-SnWO$_4$ remains the stable polymorph. The fact that β-SnWO$_4$ can be obtained as a metastable phase at room temperature is likely related to a large activation energy barrier for the β→α phase transition.

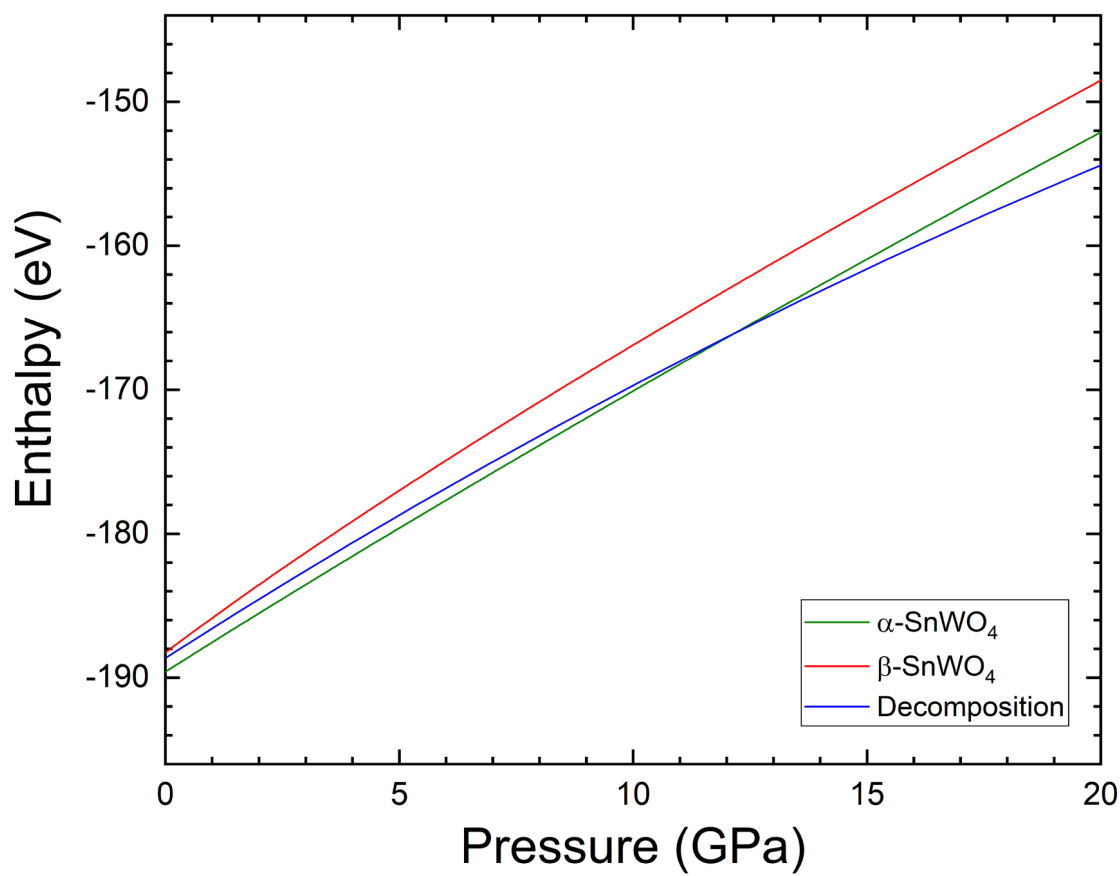

**Figure 6:** Enthalpy of α-SnWO$_4$, β-SnWO$_4$, and the chemical decomposition here proposed versus pressure.

Interestingly, as pressure increases, the enthalpy difference between α-SnWO$_4$ and β-SnWO$_4$ becomes even more pronounced in favor of the former. At 14 GPa, i.e. the decomposition pressure, the enthalpy of α-SnWO$_4$ is 4 eV smaller than that of β-SnWO$_4$. In contrast, at 13 GPa the enthalpy of the decomposition products becomes smaller than the enthalpy of α-SnWO$_4$. This implies that the energy barrier blocking the β-α transition should be at least 4 eV; a value comparable to the well-known energy barrier that hinders structural transformations in Mg$_2$SiO$_4$ olivine, which has been widely investigated due to its importance in Earth sciences. One possible reason for this kinetic barrier in SnWO$_4$ is that Sn has six-fold coordination in β-SnWO$_4$ but four-fold coordination in α-SnWO$_4$ (see Figure 1). It is well-accepted that a principal mechanism by which materials respond to an increase in pressure is through an increase in coordination [38]. Consequently, the β→α transition would contradict this principle and face significant disadvantages, as it entails the disruption of bonds, which may contribute to the energy barrier impeding the transition. In contrast, the observed decomposition does not affect the bonding of Sn and favors an increase in W coordination from tetrahedral to octahedral. Given this fact and that beyond 13 GPa the chemical decomposition of β-SnWO$_4$ becomes thermodynamically more favorable than α-SnWO$_4$, it is then not surprising to observe a chemical decomposition into more dense-packed daughter compounds, as found in this work.

Chemical decomposition has been previously observed in tungstates such as HfW$_2$O$_8$ [39]. It has been proposed that this decomposition could be triggered by mechanical or dynamical instabilities when a transition is hindered under high pressure [39]. To test the dynamical stability of β-SnWO$_4$, we calculated its phonon dispersion. We found that, within the pressure range covered by this study, all phonon branches of β-SnWO$_4$ are positive. This is illustrated in Figure 7, which shows the phonon dispersion calculated at 15.2 GPa, showing that β-SnWO$_4$ is dynamically stable.

To check the mechanical stability of β-SnWO$_4$, we calculated the elastic constants. Cubic crystals are characterized by only three elastic constants, $C_{11}$, $C_{44}$, $C_{12}$. Under hydrostatic pressure, the stability conditions are $C_{11} + 2C_{12} + P > 0$; $C_{44} - P > 0$; and $C_{11} - C_{12} + 2P > 0$ [40]. The values calculated for the elastic constants at 0 GPa are $C_{11}$ = 49.96 GPa, $C_{44}$ = 17.04 GPa, and $C_{12}$ = 13.81 GPa. At 15.2 GPa, we obtained $C_{11}$ = 173.76 GPa, $C_{44}$ = 49.83 GPa, and $C_{12}$ = 50.46 GPa. Both sets of values satisfy the stability criteria. Thus, at both 0 and 15.2 GPa, we conclude that β-SnWO$_4$ is mechanically stable. Summing up, our calculations rule out mechanical or dynamical instability as the trigger for decomposition, making the formed based on enthalpy, kinetic energy barrier and coordination theory the most rational hypothesis to explain the decomposition.

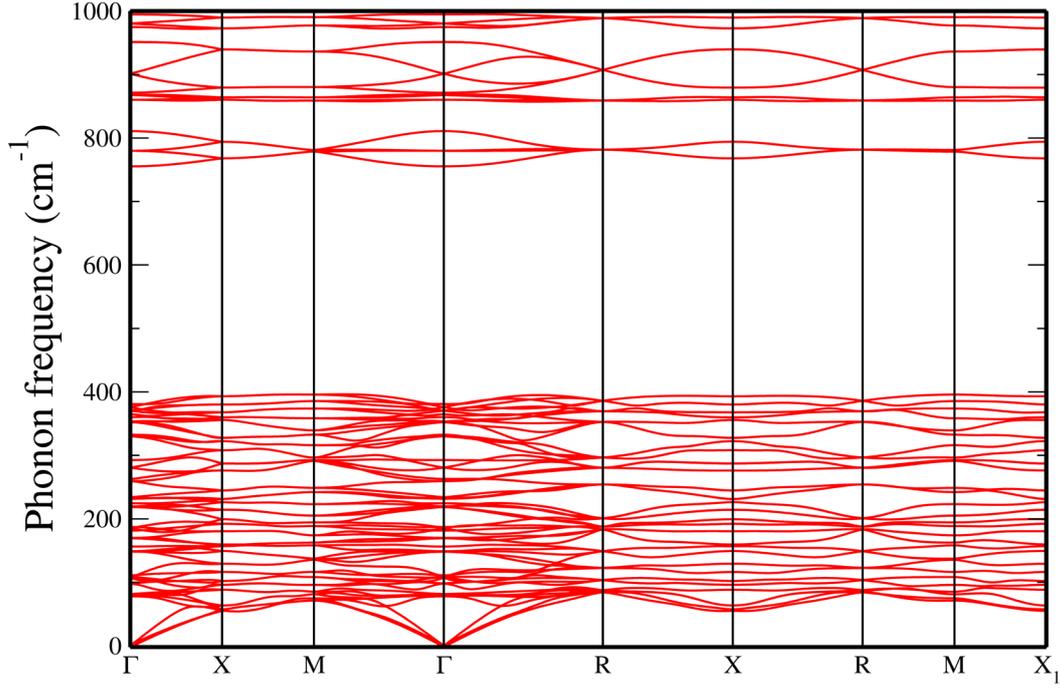

**Figure 7:** Calculated phonon dispersion of β-SnWO$_4$ at 15.2 GPa.

The polycrystalline bulk modulus (*B*) and shear modulus (*G*) can be obtained from the elastic constants. We utilized the Hill's approximation to calculate them [41]. We also derived Young's modulus (*E*) and Poisson's ratio (ν). We obtained *B* = 25.8 GPa, *G* = 17.4 GPa, *E* = 42.7 GPa, and ν = 0.225. These results indicate that the bulk modulus at zero pressure, derived from the EOS, 22.9(1.6) GPa, is similar to the bulk modulus derived from the elastic constants, 25.8 GPa, thereby validating the accuracy of our elastic calculations. The value of Young's modulus, 42.7 GPa, indicates that in β-SnWO$_4$, tensile (or compressive) stiffness under axial forces is larger than the resistance to volumetric compression. Regarding the shear modulus, its value, 17.4 GPa, suggests that β-SnWO$_4$ can be easily deformed under shear stress, making shear deformations more likely to occur than volume contraction. The fact that *B/G* = 1.48 suggests that β-SnWO$_4$ is brittle [42]. The value of Poisson's ratio, 0.225, is consistent with the fact that most solids have Poisson's ratios in the range of 0.2 to 0.3.

We also calculated the zone-zero phonon frequencies and their pressure dependence. According to group-theory analysis, β-SnWO$_4$ has 29 optical modes. Their mechanical representation is Γ = 6A + 6E + 17T. All modes are Raman-active, and the T modes are also IR-active. There are two previous experiments where the frequency of the modes was measured [43, 44]. A maximum of 24 modes were detected. Our results are compared with previous experiments in Table 4. The agreement is quite good [45], with maximum differences between experimental and theoretical wavenumbers smaller than 5%. A characteristic feature of the phonon spectrum is the existence of a phonon gap from 320 to 800 cm$^{-1}$. The high-frequency modes are due to internal vibrations of the WO$_4$ tetrahedron [44]. The low-frequency modes are associated with vibrations of the WO$_4$ tetrahedron as a rigid unit and movements of Sn atoms [44].

**Table 4:** Calculated wavenumber ($\omega_0$) for phonon frequencies at ambient pressure including mode assignment. Results are compared with previous experiments [43,44]. $a_1$, $a_2$, and $a_3$ are the coefficients of the pressure dependence of $\omega(P) = \omega_0 + a_1 P + a_2 P^2 + a_3 P^3$. When $a_3$ is not given is equal to 0 because a quadratic fit was used.

| Mode | Theory | | | | Experiments | |
|---|---|---|---|---|---|---|
| | $\omega_0$ (cm$^{-1}$) | $a_1$ (cm$^{-1}$/GPa) | $a_2$ (cm$^{-1}$/GPa$^2$) | $a_3$ (cm$^{-1}$/GPa$^3$) | $\omega_0$ Raman (cm$^{-1}$) [43] | $\omega_0$ Raman and IR (cm$^{-1}$) [44] |
| T | 40.3 | 3.07 | -0.10 | | | 36.7 |
| T | 50.5 | 5.02 | -0.13 | | | 43.3 |
| E | 54.2 | 2.47 | -0.04 | | 54 | 56.4 |
| T | 59.1 | 5.53 | -0.18 | | | 64.7 |
| A | 62.0 | 4.56 | -0.10 | | 62 | 75 |
| T | 79.5 | 4.90 | -0.01 | | 83 | 86.4 |
| E | 109.2 | 3.10 | -0.03 | | 106 | 111.3 |
| T | 124.5 | 3.73 | -0.05 | | | |
| T | 143.4 | 2.47 | 0.01 | | | |
| A | 143.8 | -0.09 | 0.12 | | | 144.5 |
| T | 154.6 | 3.86 | 0.03 | | | 158.1 |
| E | 158.4 | 8.18 | -0.27 | | | 170 |
| T | 177.1 | 9.07 | -0.17 | -0.01 | | 180.8 |
| A | 210.6 | -0.56 | 0.12 | | | 220.1 |
| A | 251.6 | 6.27 | 0.09 | | | 265.7 |
| T | 254.8 | -5.65 | 0.05 | 0.005 | | |
| E | 260.9 | -2.93 | 0.18 | | 272 | 270 |
| T | 294.0 | 2.76 | -0.03 | | | |
| E | 301.0 | 2.13 | 0.09 | | | |
| T | 307.3 | 1.03 | 0.15 | | 316 | 315 |
| T | 317.6 | 3.13 | 0.02 | | 323 | 321.5 |
| T | 320.9 | 2.73 | 0.06 | | 333 | 334.5 |
| A | 794.0 | -2.31 | -0.02 | | 769 | 798 |
| T | 797.8 | -1.72 | 0.04 | | 795 | 800.9 |
| E | 803.5 | 5.74 | -0.15 | | | 793 |
| T | 808.4 | 6.10 | -0.16 | | | 802.3 |
| T | 853.9 | 3.09 | 0 | | 899 | 850 |
| T | 947.2 | 1.94 | 0.01 | | | 900 |
| A | 949.9 | 0.77 | 0.05 | | 955 | 955.8 |

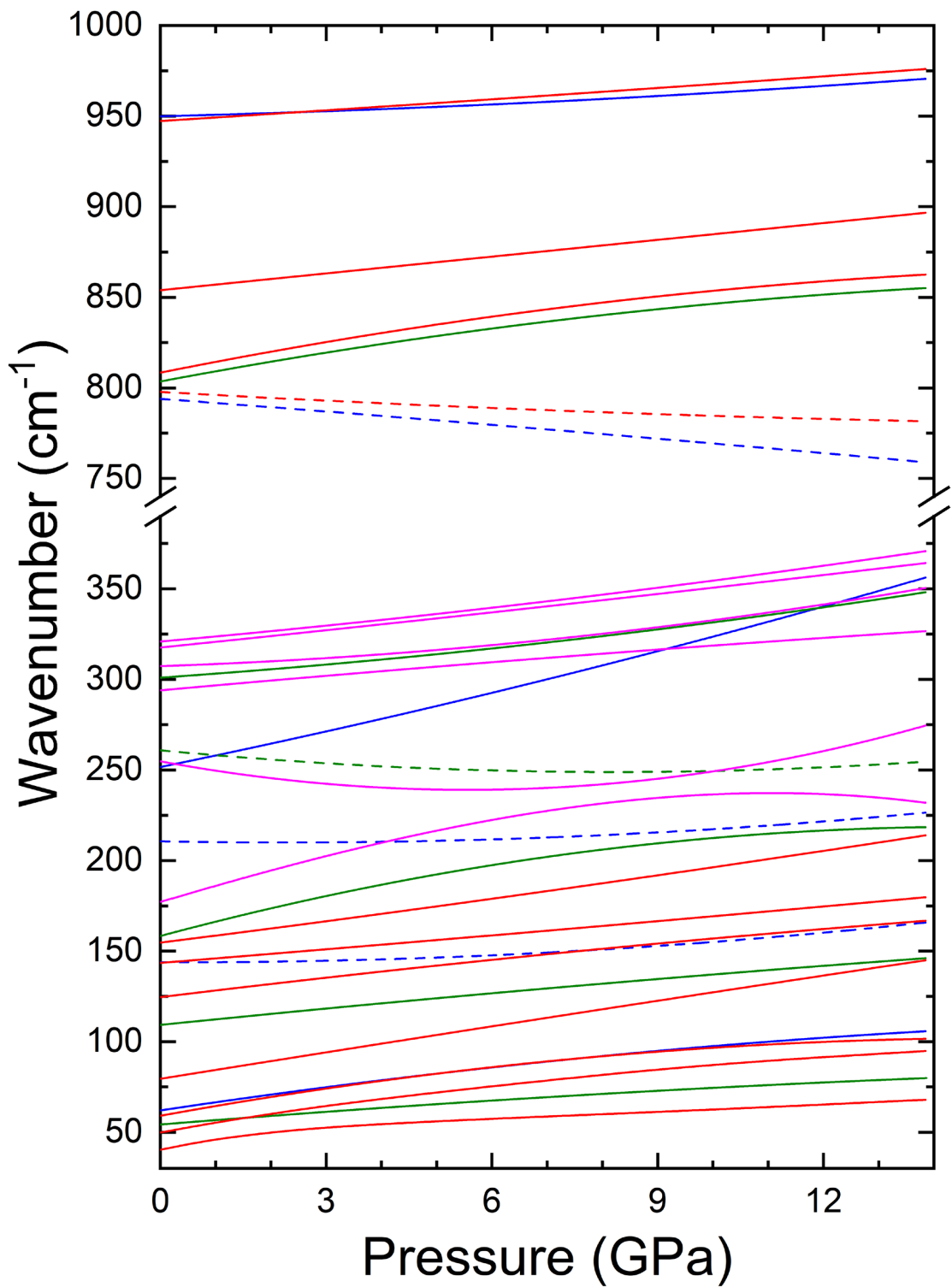

**Figure 8:** Phonon wavenumbers versus pressure. T modes are shown in red and magenta, E modes are shown in green, and A modes are shown in blue. Dashed lines are used for modes that soften under compression. The pairs of T modes shown in magenta exhibit anti-crossing.

The obtained pressure dependence of phonon wavenumbers is shown in Figure 8. Most modes follow a quadratic or linear dependence. There are only two modes exhibiting a strongly non-linear dependence. These are the T-symmetry modes with wavenumbers 177.1 and 254.8 cm$^{-1}$ at 0 GPa. The pressure dependence of the wavenumber was fitted using $\omega(P) = \omega_0 + a_1 P + a_2 P^2 + a_3 P^3$. The coefficients of the fits are given in Table 4.

The behavior of the phonons with pressure is complex, including the crossing, the anti-crossing, and the softening of several modes. Two of the high-frequency modes show significant softening. These are one A-mode (794.0 cm$^{-1}$ at 0 GPa) and one T-mode (797.8 cm$^{-1}$ at 0 GPa), represented by dashed lines. There are also two A modes (143.8 and 210.6 cm$^{-1}$ at 0 GPa) and one E mode (260.9 cm$^{-1}$ at 0 GPa) at low frequencies that soften under compression. Such a phenomenon is usually related to an enlargement of interatomic bonds [46] and might be associated with the development of structural instabilities at high pressures [47]. The phonon softening could be caused by the enlargement of some of the W-O and Sn-O distances of the WO$_4$ and SnO$_6$ polyhedra or by a charge transfer from shorter to larger bonds. Both of these facts, which will be studied in more detail below, would reduce the force constant of vibrations causing phonons to soften.

Among the modes that harden under compression, we noticed that T modes tend to show a faster increase in frequency with pressure compared to A and E modes. This causes the crossing of T modes with modes of different symmetry, as shown in Figure 8. We noticed that there are three pairs of consecutive T modes that exhibit opposite behavior. These are the pair of T modes at 320.9 and 317.6 cm$^{-1}$, 307.3 and 294.0 cm$^{-1}$, and 254.8 and 177.1 cm$^{-1}$. In each pair of modes, the mode with the lowest frequency hardens upon compression at low pressure (e.g. the mode at 177.1 cm$^{-1}$) while the mode with the highest frequency softens with pressure (e.g. the mode at 254.8 cm$^{-1}$). These modes are plotted with magenta color in Figure 8 to facilitate their identification. Due to the described phenomenon, the modes approach each other under compression. However, mode crossing cannot occur for two modes with the same symmetry [48], hence resulting in a pressure-induced phonon anti-crossing.

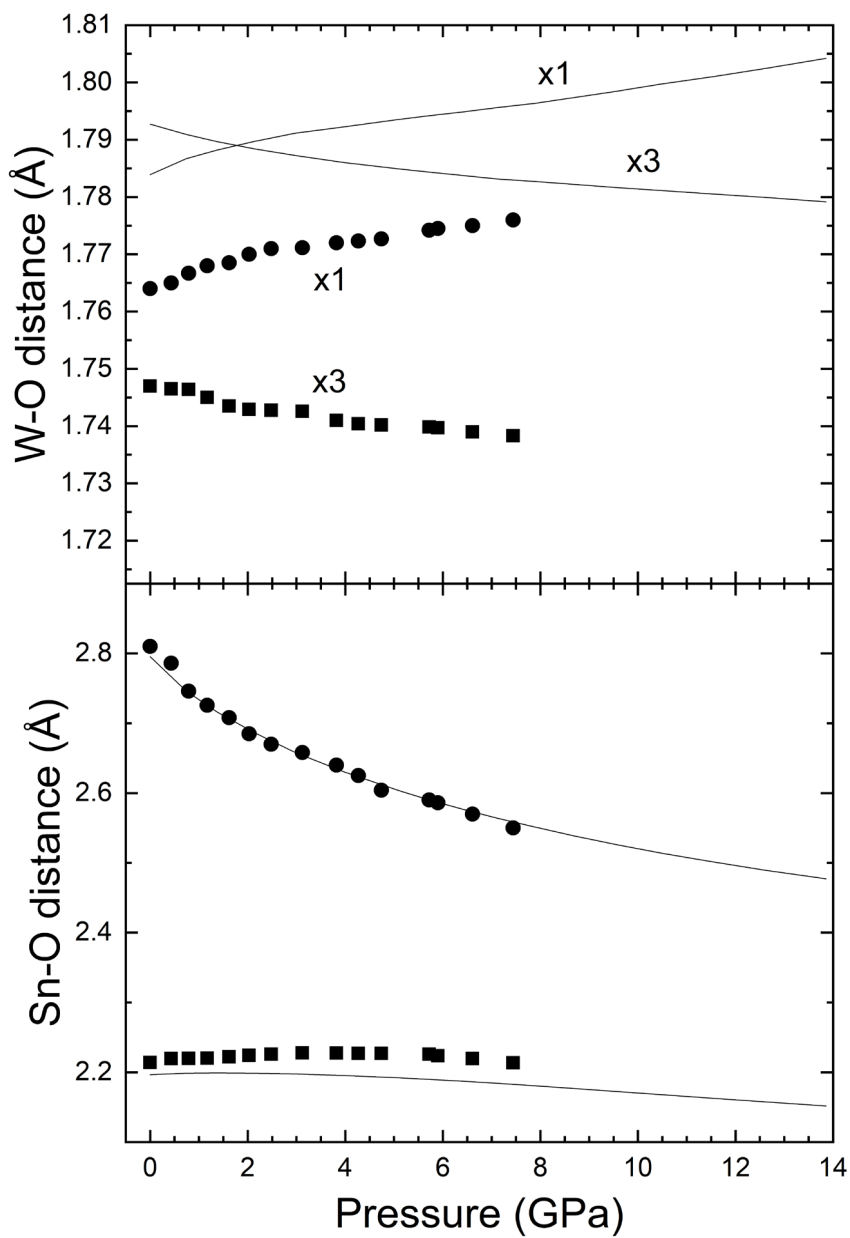

**Figure 9:** Pressure dependence of bond distances. Symbols are from experiments and lines from calculations. In the upper figure, x3 and x1 are used for W-O bond distances to distinguish the triple degenerated bonds from the fourth bond.

To conclude the present analysis, we have studied the pressure dependence of Sn-O and W-O bond distances. We obtained them from the Rietveld refinements of the XRD patterns measured up to 7.44 GPa and from DFT calculations up to 13.9 GPa, the decomposition pressure. From experiments above 7.44 GPa, we could not obtain reliable results because experiments were affected by preferred orientations. The results are shown in Figure 9.

Calculations and experiments show a similar pressure dependence for the bond distances. There is only a small difference between the W-O bond lengths around ambient pressure. There are three large bonds and one short bond according to experiments, however, calculations give one short and three longs bonds. In any case, in both cases the difference between short and long bonds is very small, of the order of 0.01 Å. In spite of this, both calculations and experiments found the non-degenerated W-O bond (marked as x1 in Figure 9) to increase with pressure. This result explains the softening of the high-frequency phonons related to internal stretching vibrations of the $WO_4$ tetrahedron. For the Sn-O bonds, we found that the long bonds rapidly decrease with pressure, while the short bonds increase up to 5 GPa and then decrease. Notice that the difference between Sn-O bond lenghts, which is of the order of 0.6 Å at 0 GPa, is of the order of 0.3 Å at the transition pressure, which means that there is a considerable restructuration of the $SnO_6$ octahedron. Most likely, this fact is responsible for the "unusual" behavior of several of the modes observed below 350 cm$^{-1}$.

## 4. Conclusions

In this study, we present findings from a high-pressure synchrotron powder X-ray diffraction investigation of β-SnWO$_4$. Our results indicate that this compound undergoes decomposition into Sn, $SnO_2$, and $WO_3$ at a pressure of 13.95 GPa, and this process is irreversible. This conclusion rules out a previous prediction [8] of a pressure-induced transition from β-SnWO$_4$ to α-SnWO$_4$. We discussed the possible reasons for decomposition. The most likely hypothesis is the existence of a large energy barrier related to the fact that Sn is six-fold coordinated in β-SnWO$_4$, but four-fold coordinated in α-SnWO$_4$. Our conclusions are supported by density-functional theory calculations, which also provide information on elastic constants and phonon frequencies. Several phonons undergo softening under pressure, while others show anti-crossing. The behavior of phonons was correlated with the behavior of bond distances. Finally, the pressure-volume equation of state for β-SnWO$_4$ was reported. This material is highly compressible. In fact, it is the most compressible known tungstate, with a bulk modulus similar to compounds with halogen bonds, which is unusual for oxides.

## Author Contributions

D. Errandonea conceived the project. S. Ferrari, D. Diaz-Anichtchenko, P. Botella, J. Ibáñez, R. Oliva, A. Kuzmin, A. Muñoz, F. Alabarse, and D. Errandonea carried out the investigation and formal analysis. All authors participated in writing and editing the manuscript. All authors have given approval to the final version of the manuscript.

## Declaration of competing interest

The authors declare that they have no known competing financial interests or personal relationships that could have appeared to influence the work reported in this paper.

## Data Availability

The data that support the findings of this study are available from the corresponding author upon reasonable request.

## Acknowledgments

The authors express their sincere gratitude for the financial assistance provided by the Spanish Ministerio de Ciencia e Innovación MCIN (DOI: 10.13039/501100011033) for project PID2022-138076NB-C41/C44. They acknowledge financial support from Generalitat Valenciana through grants CIPROM/2021/075 and MFA/2022/007. Additionally, D.E. and P.B. appreciate the support from Generalitat Valenciana via grant CIAPOS/2023/406. This research is part of the Advanced Materials program and receives support from MCIN, funded by the European Union Next Generation EU (PRTR-C17.I1), as well as from Generalitat Valenciana. The authors also thank Elettra Sincrotrone Trieste for proposal no. 20235003.

## References

[1] Qiu, W., Zhang, Y., He, G., Chen, L., Wang, K., Wang, Q., Li, W., Liu, Y., Li, J. Two-Dimensional Long-Plate $SnWO_4$ Photoanode Exposed Active Facets for Enhanced Solar Water Splitting. ACS Applied Energy Materials 2022, 5, 11732-11739. DOI: 10.1021/acsaem.2c02235

[2] Dan, M., Cheng, M., Gao, H., Zheng, H., Feng, C. Synthesis and electrochemical properties of $SnWO_4$, J. Nanoscie. Nanotech. 2014, 14, 2395–2399. DOI: 10.1166/jnn.2014.8497

[3] Kölbach, M., Jordão Pereira, I., Harbauer, K., Plate, P., Höflich, K., Berglund, S.P., Friedrich, D., van de Krol, R., Abdi, F.F. Revealing the Performance-Limiting Factors in α-SnWO$_4$ Photoanodes for Solar Water Splitting. Chemistry of Materials 2018, 30, 8322-8331. DOI: 10.1021/acs.chemmater.8b03883

[4] Solis, J.L., Lantto, V. Gas-sensing properties of different α-SnWO$_4$-based thick films, Phys. Scr. 1997, T67, 281–285. DOI: 10.1088/0031-8949/1997/T69/059

[5] Halasyamani, P. S. Asymmetric Cation Coordination in Oxide Materials: Influence of Lone-Pair Cations on the Intra-octahedral Distortion in d$^0$ Transition Metals. Chem. Mater. 2004, 16, 3586–3592. DOI: 10.1021/cm049297g

[6] Jeitschko, W., Sleight, A.W. Synthesis, Properties and Crystal Structure of β-SnWO$_4$. Acta Crystallogr. B 1972, 28, 3174-3178. DOI: 10.1107/S056774087200768X

[7] Jeitschko W, Sleight, A.W. α-Stannous Tungstate: Properties, Crystal Structure and Relationship to Ferroeleetric SbTaO$_4$ Type Compounds. Acta Crystallogr. 1974. B30, 2088–2094. DOI: 10.1107/S0567740874006534

[8] Gomes, E.O., Gouveia, A.F., Gracia, L., Lobato, A., Recio, J.M., Andrés, J. A Chemical-Pressure-Induced Phase Transition Controlled by Lone Electron Pair Activity. J. Phys. Chem. Letters 2022, 13, 9883-9888. DOI: 10.1021/acs.jpclett.2c02582

[9] Kuzmin, A., Anspoks, A., Kalinko, A., Timoshenko, J., Kalendarev, R. External pressure and composition effects on the atomic and electronic structure of SnWO$_4$, Solar Energy Materials and Solar Cells 2015, 143, 627-634. DOI: 10.1016/j.solmat.2014.12.003.

[10] Mao, H.K., Chen, X.J., Ding, Y., Li, B., Wang, L. Solids, liquids, and gases under high pressure- Rev. Mod. Phys. 2018, 90, 015007. DOI: 10.1103/RevModPhys.90.015007

[11] Errandonea, D., Osman, H., Turnbull, R., Diaz-Anichtchenko, D., Liang, A., Sanchez-Martin, J., Popescu, C., Jiang, D., Song, H., Wang, Y., Manjon, F.J. Pressure- induced hypercoordination of iodine and dimerization of I$_2$O$_6$H in strontium di- iodate hydrogen-iodate (Sr(IO$_3$)$_2$HIO$_3$), Mat. Today Adv 2024, 22, 100495. DOI: 10.1016/j.mtadv.2024.100495

[12] Diaz-Anichtchenko, D., Ibáñez, J., Botella, P., Oliva, R., Kuzmin, A., Wang, L., Li, Y., Muñoz, A., Alabarse, F., Errandonea, D. Identification of the high-pressure phases of α-SnWO$_4$ combining x-ray diffraction and crystal structure prediction, Physica B 2025, 696, 416666. DOI: 10.1016/j.physb.2024.416666.

[13] Klotz, S., Chervin, J.C., Munsch, P., Le Marchand, G. Hydrostatic limits of 11 pressure transmitting media, J. Phys. D Appl. Phys. 2009, 42, 075413. DOI: 10.1088/0022-3727/42/7/075413

[14] Díaz-Anichtchenko, D., Santamaria-Perez, D., Marqueño, T., Pellicer-Porres, J., Ruiz-Fuertes, J., Ribes, R., Ibañez, J., Achary, S.N., Popescu, C., Errandonea, D. Comparative study of the high-pressure behavior of ZnV$_2$O$_6$, Zn$_2$V$_2$O$_7$, and Zn$_3$V$_2$O$_8$, J. Alloys Compd. 2020, 837, 155505. DOI: 10.1016/j.jallcom.2020.155505

[15] Dewaele, A., Loubeyre, P., Mezouar, M. Equations of state of six metals above 94 GPa, Phys. Rev. B, 2004, 70, 094112. DOI: 10.1103/PhysRevB.70.094112

[16] Prescher, C., Prakapenka, V.B. DIOPTAS: a program for reduction of two- dimensional X-ray diffraction data and data exploration, High Pres. Res. 2015, 35, 223–230. DOI: 10.1080/08957959.2015.1059835

[17] Lutterotti, L. Maud: A Rietveld Analysis Program Designed for the Internet and Experiment Integration. Acta Crystallographica A 2000, 56, S54. DOI: 10.1107/S0108767300021954

[18] Jones, R.O. Density functional theory: its origins, rise to prominence, and future. Rev. Mod. Phys. 2015, 87, 897–923. DOI: 10.1103/RevModPhys.87.897

[19] Blöchl, P.E. Projector augmented-wave method. Phys. Rev. B 1994, 50, 17953 – 17979. DOI: 10.1103/PhysRevB.50.17953

[20] Kresse, G., Hafner, J., Ab initio molecular dynamics for liquid metals, Phys. Rev. B 1993, 47, 558–561. DOI: 10.1103/PhysRevB.47.558

[21] Perdew, J.P., Ruzsinsky, A., Csonka, G.I., Vydrov, O.A., Scuseria, G.E., Constantin, L.A., Zhou, X., Burke, K. Restoring the density-gradient expansion for exchange in solids and surfaces. Phys. Rev. Lett. 2008, 100, 136406. DOI: 10.1103/PhysRevLett.100.136406

[22] Monkhorst, H.J., Pack, J.D. Special points for Brillouin-zone integrations. Phys. Rev. B 1976, 13, 5188–5192. DOI: 10.1103/PhysRevB.13.5188

[23] Togo, A. First-principles Phonon Calculations with Phonopy and Phono3py" J. Phys. Soc. Jpn. 2023, 92, 012001. DOI: 10.7566/JPSJ.92.012001

[24] Le Page, Y., Saxe, P. Symmetry-general least-squares extraction of elastic data for strained materials from ab initio calculations of stress. Phys. Rev. B 2002, 65, 104104. DOI: 10.1103/PhysRevB.65.104104

[25] Birch, F. Finite elastic strain of cubic crystals. Phys. Rev. 1947, 71, 809-824. DOI: 10.1103/PhysRev.71.809

[26] Errandonea, D., Ferrer-Roca, C., Martínez-Garcia, D., Segura, A., Gomis, O., Muñoz, A., Rodríguez-Hernández, P., López-Solano, J., Alconchel, S., Sapiña, F., High-pressure x-ray diffraction and ab initio study of $Ni_2Mo_3N$, $Pd_2Mo_3N$, $Pt_2Mo_3N$, $Co_3Mo_3N$, and $Fe_3Mo_3N$: two families of ultra-incompressible bimetallic interstitial nitrides. Phys. Rev. B 2010, 82, 174105. DOI: 10.1103/PhysRevB.82.174105

[27] Errandonea, D., Manjón, F.J., Pressure effects on the structural and electronic properties of $ABX_4$ scintillating crystals Progress in Materials Science 2008, 53, 711-773. DOI: 10.1016/j.pmatsci.2008.02.001

[28] Errandonea, D., Ruiz-Fuertes, J. A Brief Review of the Effects of Pressure on Wolframite-Type Oxides. Crystals 2018, 8, 71. DOI: 10.3390/cryst8020071

[29] Liang, A., Turnbull, R., Errandonea, D. A review on the advancements in the characterization of the high-pressure properties of iodates. Prog. Mater. Sci. 2023, 136, 101092. DOI: 10.1016/j.pmatsci.2023.101092

[30] Liang, A., Gonzalez-Platas, J., Turnbull, R., Popescu, C., Fernandez-Guillen, I., Abargues, R., Boix, P.P., Shi, L.T., Errandonea, D. Reassigning the Pressure-Induced Phase

Transitions of Methylammonium Lead Bromide Perovskite. J. Am. Chem. Soc. 2022, 144, 20099-20108. DOI: 10.1021/jacs.2c09457

[31] Panchal, V.; Pampillo, L.; Ferrari, S.; Bilovol, V.; Popescu, C.; Errandonea, D. Pressure-Induced Structural Phase Transition of Co-Doped $SnO_2$ Nanocrystals. Crystals 2023, 13, 900. DOI: 10.3390/cryst13060900

[32] Bouvier, P., Crichton, W.A., Boulova, M., Lucazeau, G. X-ray diffraction study of $WO_3$ at high pressure, J Phys: Condens Matter 2002, 14, 6605-6617. DOI: DOI 10.1088/0953-8984/14/26/301

[33] Desgreniers, S., Vohra, Y.K., Ruoff, Al.L. Tin at high pressure: An energy-dispersive x-ray-diffraction study to 120 GPa, Phys. Rev. B 1989, 39, 10359 - 10361. DOI: 10.1103/PhysRevB.39.10359

[34] Bozheyev, F., Fengler, S., Kollmann, J., Abou-Ras, D., Scharnagl, N., Schieda, M. Influence of $SnWO_4$, $SnW_3O_9$, and $WO_3$ Phases in Tin Tungstate Films on Photoelectrochemical Water Oxidation. ACS Applied Materials & Interfaces 2024, 16, 48565-48575. DOI: 10.1021/acsami.4c09713

[35] Chen, X., Lou, H., Zeng, Z., Cheng, B., Zhang, X., Liu, Y., Xu, D., Yang, K., Zeng, Q. Structural transitions of 4:1 methanol–ethanol mixture and silicone oil under high pressure. Matter Radiat. Extremes 2021, 6, 038402. DOI: 10.1063/5.0044893

[36] Karaca, E., Santamaria-Perez, D., Otero-de-la-Roza, A., Oliva, R., Rao, K.S., Achary, S.N., Popescu, C., Errandonea, D, Pressure-induced decomposition of $Bi_{14}WO_{24}$, Results in Physics 2025, 70, 108170. DOI: 10.1016/j.rinp.2025.108170.

[37] Mahendran, S., Carrez, P., Cordier, P. On the glide of [100] dislocation and the origin of 'pencil glide' in $Mg_2SiO_4$ olivine. Philosophical Magazine 2019, 99, 2751–2769. DOI: 10.1080/14786435.2019.1638530

[38] Serghiou, G., Boehler, R., Chopelas, A. Four to sixfold coordination changes in crystalline silicates at high pressure. High Pressure Research 2003, 23, 49–53. DOI: 10.1080/0895795031000109689

[39] Sakuntala, T., Rao, R., Garg, A.B., Achary, S.N., Tyagi, A.K. Amorphization-decomposition behavior of $HgW_2O_8$ at high pressure. J. Appl. Phys. 2008, 104, 063506. DOI: 10.1063/1.2973193

[40] Grimvall, G., Magyari-Köpe, B., Ozoliņš, V., Persson, K.A., Lattice instabilities in metallic elements. Rev. Mod. Phys. 2012, 84, 945-986. DOI: 10.1103/RevModPhys.84.945

[41] Hill, R. The Elastic Behaviour of a Crystalline Aggregate. Proc. Phys. Soc., London, Sect. A 1952, 65, 349-362, DOI: 10.1088/0370-1298/65/5/307

[42] Senkov, O.N., Miracle, D.B. Generalization of intrinsic ductile-to-brittle criteria by Pugh and Pettifor for materials with a cubic crystal structure. Sci Rep 2012, 11, 4531. DOI: 10.1038/s41598-021-83953-z

[43] Solis, J.L., Frantti, J., Lantto, V., Häggström, L., Wikner, M. Characterization of phase structures in semiconducting $SnWO_4$ powders by Mössbauer and Raman spectroscopies. Phys. Rev. B 1998, 57, 13491-13500. DOI: 10.1103/PhysRevB.57.13491

[44] Wojcik, J., Calvayrac, F., Goutenoire, F., Mhadhbi, N., Corbel, G., Lacorre, P., Bulou, A. Lattice Dynamics of β-SnWO$_4$: Experimental and Ab Initio Calculations. J. Phys. Chem. C 2013, 117, 5301-5313. DOI: 10.1021/jp3099126

[45] Errandonea, D., Muñoz, A., Rodríguez-Hernández, P., Gomis, O., Achary, S.N., Popescu, C., Patwe, S.J., Tyagi, A.K. High-Pressure Crystal Structure, Lattice Vibrations, and Band Structure of BiSbO$_4$. Inorganic Chemistry 2016, 55, 4958-4969. DOI: 10.1021/acs.inorgchem.6b00503

[46] Diaz-Anichtchenko, D., Aviles-Coronado, J.E., López-Moreno, S., Turnbull, R., Manjón, F.J., Popescu, C., Errandonea, D. Electronic, Vibrational, and Structural Properties of the Natural Mineral Ferberite (FeWO$_4$): A High-Pressure Study. Inorg. Chem. 2024, 63, 15, 6898–6908. DOI: 10.1021/acs.inorgchem.4c00345

[47] Dove, M.T. Theory of displacive phase transitions in minerals. American Mineralogist 1997, 82, 213–244. DOI: 10.2138/am-1997-3-401

[48] Hsu, L., McCluskey, M.D., Lindström, J.L. Resonant Interaction between Localized and Extended Vibrational Modes in Si:$^{18}$O under Pressure, Phys. Rev. Lett. 2003, 90, 095505. DOI: 10.1103/PhysRevLett.90.095505